\documentclass[final]{aa}
\usepackage{graphicx}
\usepackage{txfonts}
\begin{document}
\title{Spectroscopic analysis of tremendous--outburst--nova candidates
\thanks{Based on observations collected at the European Southern Observatory, 
        La Silla, Chile}}
\titlerunning{Spectroscopic analysis of TON candidates}
\author{L. Schmidtobreick \inst{1}
        \and
        C. Tappert \inst{2}
        \and
        A. Bianchini \inst{3}
        \and
        R.\,E. Mennickent \inst{4}}
\offprints{Linda Schmidtobreick, \email{lschmidt@eso.org}}
\institute{European Southern Observatory, Casilla 19001, Santiago 19, Chile.
           \and 
           Departamento de Astronom\'{\i}a y Astrof\'{\i}sica, 
           Pontificia Universidad Cat\'olica, Casilla 306, Santiago 22, Chile
           \and
           Dipartimento di Astronomia, Universit\`a di Padova,
           Vicolo dell'Osservatorio 2, I-35122, Padova, Italy
           \and
           Grupo de Astronom\'{\i}a, Universidad de Concepci\'on,
           Casilla 160--C, Concepci\'on, Chile}
\date{Received xxx xxx, xxx; accepted xxx xxx, xxx}
\abstract{
In the course of a long-term project investigating classical novae with
large outburst
amplitudes, we have performed optical spectroscopy
of several old--nova candidates. We here present the
spectra of the candidates 
V630\,Sgr, XX\,Tau, CQ\,Vel, V842\,Cen, and V529\,Ori, that hitherto lacked such
classification.
 While the first four show typical spectra for cataclysmic variables, and can 
thus be identified as such, V529\,Ori is probably misclassified. 
Of special interest are the two systems XX\,Tau and V842\,Cen, which show all 
signs for being low mass transfer systems. As such they can be used to judge
the evolution scenarios for novae. 
In particular, given the rather young age of their outburst, 
it appears more likely that these systems are not on their way into hibernation,
(i.e., cutting off mass transfer for a longer period of time), 
but are simply settling down towards their original configuration of
comparatively low, but steady, mass transfer, such as for dwarf novae.
   }
\maketitle

\section{Introduction}
Novae are a subclass of cataclysmic variable stars (CVs) 
(see Warner \cite{warn95} for a general introduction on these objects)
and hence are interacting binary systems, which consist of a white dwarf 
primary accreting mass from a main sequence like late type secondary star.
In the subclass of novae the such gradually increasing hydrogen--rich
material on the surface of the white dwarf, has at one point reached a 
critical mass which resulted in a thermonuclear runaway, the nova explosion.

The evolution of CVs generally represents an evolution towards shorter periods
due to continuous loss of angular momentum by magnetic braking and
gravitational radiation. Near an orbital period of 78\,min, however, the
secondary star becomes degenerate, and the loss of mass now leads to an
increase of the separation and thus an increasing orbital period. 
At this period minimum the thermal timescale of the secondary becomes
longer than the timescale of angular momentum loss by gravitational
radiation, leading to a drastic increase of the evolutionary lifetime
of the CV near this point.
Evolutionary models predict that the vast majority of CVs should have 
already evolved beyond
that point, yielding a concentration of systems close to the period minimum
(Stehle et al. \cite{steh+96}). 
This, however, is in sharp contrast to the observed period distribution
(Ritter \& Kolb \cite{ritt+98}).
It can be in part understood as a possible observational bias, as these
systems are supposed to inhabit very
low mass--transfer rates and thus to be intrinsically very faint (Stehle et
al. \cite{steh+97}).
Dwarf novae
that show very large outburst amplitudes, so--called TOADs (tremendous
outburst amplitude dwarf novae), are therefore
generally seen as good candidates for being evolved CVs 
(Howell et al. \cite{howe+97}).

The outburst mechanism of a classical nova is a physically different
phenomenon than the disc outburst of a dwarf nova. Still, it is reasonable
to suspect a similar correlation between the outburst amplitude --
measured as the difference between the outburst peak magnitude and the
quiescence magnitude of the post nova -- and the mass transfer rate of the 
post nova. 
It can be assumed that the
absolute magnitude of a nova explosion differs only slightly for different
systems, as it depends mainly on the mass of the white dwarf 
(Livio \cite{livi92}), 
the latter depending only weakly on the orbital period (Ritter \& Kolb 1998).
Recent
studies show that accretion discs may reform within months after a nova
outburst (Retter et al. \cite{rett+98}).
Therefore, similar to the case of dwarf novae, 
an intrinsically faint system would be indicated by an
unusual large outburst amplitude.
The low intrinsic brightness might be either 
due to the CV being at high inclination
(see Warner \cite{warn87} for a detailed analysis of the 
inclination/outburst amplitude dependency) 
or because the post nova inhabits a faint
accretion disc with a low mass--transfer rate. 
We therefore use this phenomenological approach to examine the nova
population for a possible subclass of low mass transfer systems.

The existence of such a tremendous outburst nova (TON) population
could strengthen the so--called
`hibernation scenario', which proposes an evolutionary bond between several CV
subclasses (Shara et al. \cite{shar+86}). The basic statements of this
scenario are a) that all CVs undergo nova outbursts, b) that they show more
than one such outburst in their lifetime, with recurrence times $> 10^4$
years, and c) that they vanish into a state of hibernation between these 
outbursts. Other subgroups of CVs, e.g. dwarf novae or nova--likes, would thus
represent novae between two outbursts. Theoretical models indeed show that nova
outbursts should be possible even after a stage of very low mass--transfer
(Prialnik \& Shara \cite{pria+86}). Observational evidence, however, is
still missing.

We have started a project to examine the nova population for possible
TONs. One part consists in the recovery and
identification of lost old novae and the determination of
their quiescent magnitude to determine the outburst amplitude.
In this paper, we present the spectroscopic analysis for 
five objects which have been reported as recovered novae but with so far 
uncertain classification.
\section{Observation and data reduction}
\begin{table}[bt]
\caption{\label{obstab} Summary of the observational details.} 
\begin{tabular}{l l l l l}
\hline
 \noalign{\smallskip}
Object & Date & Instrument & Grism/Slit & $t_{\rm Exp}$ [s] \\
 \noalign{\smallskip}
\hline
 \noalign{\smallskip}
V630\,Sgr & 2001-07-16 & DFOSC/1.54D & G4/1.5" & 2400 \\
XX\,Tau   & 2003-01-06 & EFOSC2/3.6 & G11/2.0" & 2400 \\
V529\,Ori & 2003-01-07 & EFOSC2/3.6 & G11/2.0" & 1200 \\
CQ\,Vel   & 2003-01-07 & EFOSC2/3.6 & G11/2.0" & 4800 \\
V842\,Cen & 2003-05-12 & EMMI/NTT   & G3/1.0"  & 9300 \\
 \noalign{\smallskip}
\hline
\end{tabular}
\end{table}

The observations have been performed in the years from 2001 to 2003 at La Silla
Observatory, Chile, using DFOSC at the 1.54m Danish telescope, EMMI at the 
3.5m New Technology Telescope, or EFOSC at the 3.6m telescope.
The details of these observations are given 
in table \ref{obstab}.

Standard reduction of the data has been done using IRAF.
The BIAS has been subtracted and the data have been divided by a flat field,
which was normalised by fitting Chebyshev functions of high order 
to remove the detector specific spectral response. 
The spectra have been optimally extracted (Horne \cite{horn86}).
Wavelength calibration yielded a final FWHM resolution of  1.0\,nm for DFOSC,
0.84\,nm for EMMI, and 1.2\,nm for the EFOSC data.

Rough flux calibration has been performed with respect to spectrophotometric 
standards EG\,274 (DFOSC),
LTT\,4363 (EMMI), and LTT\,3864 (EFOSC).
Since the
nights have not been photometric, the absolute flux values have an uncertainty
of at least 20\%. 
Relative fluxes, used to compare different parts of the spectrum 
(see Table \ref{line_tab} for details), are naturally more accurate.

Since no information on the individual interstellar extinction are
available for any of these novae, we present the spectra as observed, 
i.e. without reddening correction. 
The reddening issue is discussed in more detail for the individual 
systems when appropriate.

\section{V630 Sagittarii}
\begin{figure}
\rotatebox{-90}{\resizebox{!}{8.9cm}{\includegraphics{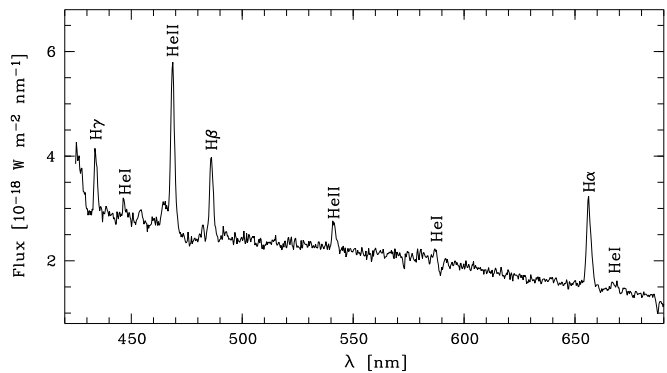}}}
\caption{\label{v630s}The spectrum of V630\,Sgr is dominated by Balmer and
He\,{\sc ii} lines in emission.}
\end{figure}

V630 Sgr has been detected in outburst in 1936 
(Okabayashi \cite{okab36}). While all older sources (Parenago \cite{pare49}, 
Gaposchkin \cite{gapo55}; Kukarkin et al. \cite{kuka+69}) list the object
with a visual magnitude of $ v = 4.^{\rm m}0$ at maximum, 
suddenly a value of
$ 1.^{\rm m}6$ appears in the fourth 
edition of the GCVS (Kholopov et al. \cite{khol+83}). 
The only additional reference with respect to previous editions of this 
  catalogue is the paper by 
Gaposchkin \cite{gapo55}, but Gaposchkin's light curve also gives
$ v = 4.^{\rm m}0$ at maximum. There it is also mentioned that on a 
photographic plate
taken the day before the nova discovery, the object was still below 
$13^{\rm m}$, which makes it very unlikely that the maximum has been missed. 
Parenago (\cite{pare49}) especially argues with the
shape of their light curve that the maximum of the nova has not been missed
but is actually around $4.^{\rm m}0$.
For these reasons we believe that the values have been changed by mistake 
between edition 3 and 4 of the GCVS, and that the now listed maximum value 
$1.^{\rm m}6$ (Downes et al. \cite{down+01}) is actually wrong, the real 
maximum value being $v = 4.^{\rm m}0$.

With $t_3 = 6\rm d$ V630\,Sgr is among the fastest novae ever observed,
Duerbeck (\cite{duer81}) classified
it as A (fast decline without major disturbances). He also used the light curve
to derive its distance $d \rm = 2000\,pc$. Shafter (\cite{shaf97}), who
later estimated the value to be $d \rm = 600\,pc$ had used the wrong maximum 
value. Using instead the value of $4.^{\rm m}0$, his method yields
 $d \rm = 1820\,pc$, well in agreement with Duerbeck.
From the short decay time 
and large amplitude (although derived from the $1.^{\rm m}6$ maximum value)
of V630\,Sgr, Diaz \& Steiner (\cite{diaz+91}) concluded that the 
nova might be of magnetic type.
Harrison \& Gehrz (\cite{harr+94}) have observed V630\,Sgr in four IR bands
with IRAS but obtained no detection in any of them. 
Recent high--speed 
photometry by Woudt \& Warner (\cite{woud+01}) revealed a shallow eclipse as
well as permanent superhumps for this nova remnant. They determined the
orbital period $P_{\rm orb}= 2.83$\,h and the superhump period 
$P_{\rm sh} = 2.98$\,h.

The spectrum of V630\,Sgr (see Fig. \ref{v630s}) is dominated by
strong Balmer and He\,{\sc ii} lines in emission and thus confirms the nova 
recovery. The emission lines show widths between 1.5\,nm and 2.4\,nm (see 
Table \ref{line_tab}), which 
compute to an average projected rotation velocity of 
$\rm 1080\pm 50\,km\,s^{-1}$.
Note also that, in spite of the
high inclination of the system, the emission lines are rather narrow and
not double--peaked at our resolution of $\sim$1\,nm. This might indicate that
the disc is optically thick in the hydrogen lines.

\begin{figure}
\rotatebox{-90}{\resizebox{!}{8.9cm}{\includegraphics{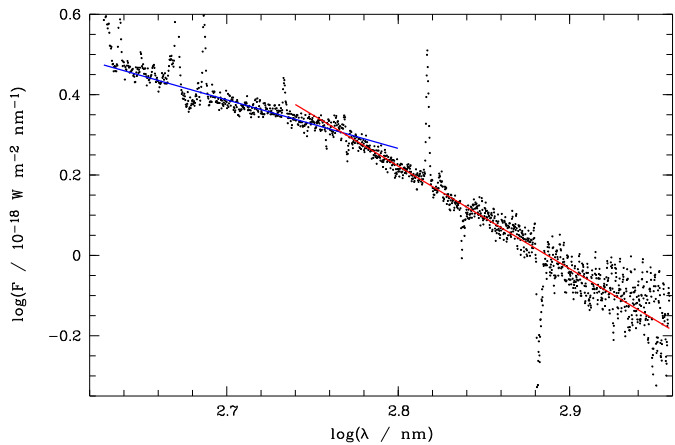}}}
\caption{\label{v630_slope}With the double logarithmic scaling it becomes
obvious that two slopes are needed to fit the continuum of V630\,Sgr.
While the red part of the continuum yields a value of $\alpha = -2.5$,
the blue part is fitted with $\alpha = -1.2$.}
\end{figure}

The possibility of a hot, optically thick, disc is also supported by the
$F = \lambda^{\alpha}$ description of the continuum slope requiring a rather
high value $\alpha = -2.5$. In this context we note that the continuum of
V630\,Sgr cannot be described by a single power law, but that it consists
of two different slopes (Fig. 2).

The assumption of a hot accretion disc with $\alpha = -2.5$ fits only the
redder part of the spectrum down to a wavelength of 582\,nm and a
corresponding temperature $T = 25000\rm K$. For shorter wavelengths and thus
higher temperatures we find a different slope of $\alpha = -1.2$.

This flatter slope reflects
the absence of an accretion disc at short radii, and instead the presence 
of a different continuum emitting region, which might be associated with the
suspected magnetic accretion of this system or an  optically thin region
in the inner part of the accretion disc. Time--resolved spectroscopy and
a detailed investigation of the emission distribution in the accretion disc
of this system is necessary to clarify these issues.
\section{XX Tauri}
\begin{figure}
\rotatebox{-90}{\resizebox{!}{8.9cm}{\includegraphics{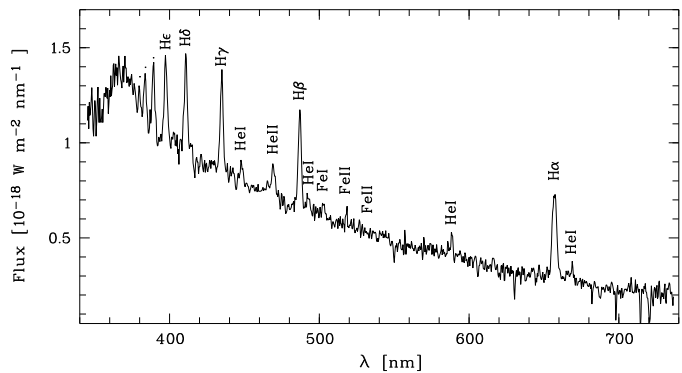}}}
\caption{\label{xxtaus}Apart from the Balmer series in emission, the spectrum 
of XX\,Tau shows various low ionisation lines like He\,I and Fe\,II.
He\,II is only present at 469\,nm.}
\end{figure}

XX\,Tau has been discovered in November 1927 by Schwassmann \& Wachmann
(\cite{schw+28}). Analyses of older Harvard plates give the first light curve
and show that the maximum at 5.9p was reached on October 1, 1927 
(Cannon \cite{cann27}).
Several minima during the early decline (Payne--Gaposchkin \cite{payn57})
suggest the formation of dust in this
nova outburst. The same feature is addressed by Duerbeck's (\cite{duer81}) 
classification as Cb of which class FH\,Ser is the prototype. 
He also gives $t_3 = 42$\,d, which places XX\,Tau among 
the moderately fast novae. 

XX\,Tau has been recovered by Cohen (\cite{cohe85}) via H$\alpha$ photometry. 
She has also been able to spatially resolve the shell around the nova in 
H$\alpha$ and R and gives its radius as 2.2\,arcsec. From the expansion 
parallax that she derived, the distance of XX\,Tau is determined as 3.5\,kpc
(Shafter \cite{shaf97}). Downes \& Duerbeck (\cite{down+00}) determined the
interstellar extinction towards XX\,Tau as $A_V = 1.26\pm 0.57$.
The presence of hot dust is confirmed by the IRAS data
of Harrison \& Gehrz (\cite{harr+94}) who detected the nova at 12\,$\mu$m
and 25\,$\mu$m.  
No detection has been accomplished in the 2mass second incremental data release 
(Hoard et al. \cite{hoar+02}) though.

The spectrum of XX\,Tau is given in Fig. \ref{xxtaus}. It is dominated by
emission lines and hence confirms Cohen's recovery. However, the presence 
of the Balmer lines down to H\,11 and the strength of He\,I compared
to He\,II gives the object the appearance of a typical dwarf nova
rather than an old classical nova. We tentatively conclude therefore that
XX\,Tau represents an old nova with a low mass transfer rate, which has 
sufficiently cooled down to look like a 'normal' dwarf nova. Since our
spectrum furthermore does not show any spectral signatures of the secondary
star, we expect XX\,Tau to have a comparatively short orbital period.

The line widths as given in 
Tab.\,\ref{line_tab} compute to 
a velocity of
$\rm 1550 \pm 40\,km\,s^{-1}$. 
Assuming this to be the radial projection of the rotational velocity,
these moderately high values suggest that
XX\,Tau is seen at 
sufficiently high inclination to make it an interesting system for time resolved
follow--up observations.

\section{V529 Orionis}
\begin{figure}
\rotatebox{-90}{\resizebox{!}{8.9cm}{\includegraphics{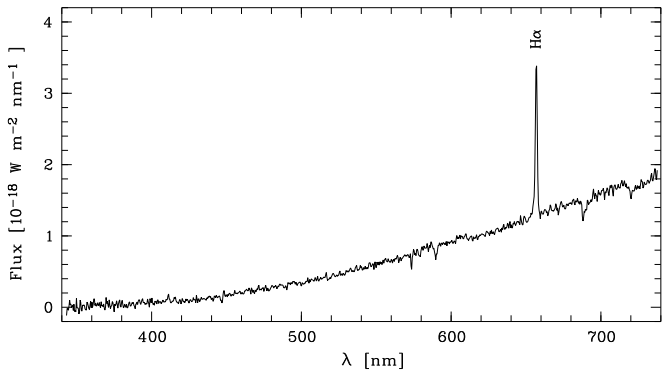}}}
\caption{\label{v528oris}The spectrum of V529\,Ori is dominated by an extremely
red continuum and the strong H$\alpha$ line in emission.}
\end{figure}

V529\,Ori is the oldest nova in our sample, and has been discovered by
J. Hevelius on March 28, 1678 while he was observing a lunar occultation 
of $\chi ^1$\,Ori. V529\,Ori or 48\,Ori, as Hevelius called it, "followed 
the path of $\chi ^1$\,Ori with respect to the moon and was occulted
between $\rm 9^h16^m$ and  $\rm 10^h29^m$" (Hevelius \cite{hev1679}). 
From these observations,
Ashworth (\cite{ashw81}) recalculated the position of V529\,Ori using
modern values for the coordinates of $\chi ^1$\,Ori. He also showed
that a supposed later observation of this object in 1750 is actually just a 
quotation of Hevelius' catalogue and thus concluded that V529\,Ori has not been
observed as recurrent nova as claimed before. 

Several attempts to recover this old nova failed. 
The here observed candidate has been proposed by Robertson et al. 
(\cite{robe+00})
on the basis of showing variability and H$\alpha$ emission.
Hoard et al. (\cite{hoar+02}) determined the NIR colours
$H-K = +0.96$ and $J-H = +1.33$ of this candidate which turn out to be
much redder
than all other CVs in their catalogue and resemble more those of 
symbiotic stars. 

The spectrum of V529\,Ori shows indeed a strong and narrow H$\alpha$ line 
on a very red continuum, as shown already by Robertson et al. (\cite{robe+00}).
From our S/N we get upper limits on the strength of H$\beta$ for the 
equivalent width $W < 0.25$\,nm and for the line flux 
$F < 0.07\cdot 10^{-18} \rm W m^{-2}$. Comparing these values with the 
strength of H$\alpha$ (see Tab. \ref{line_tab}) yields a Balmer decrement 
$\rm H\alpha / H\beta > 50$ while 7 is about the maximum for any Balmer 
emission present in CVs (Williams \cite{will91}). 
Exploring the possibility that this extreme ratio is caused by interstellar
reddening, we find that a very high value of $E_{B-V} = 1.5$ is required to
transform the observed slope to a flat continuum. Furthermore, the ratio
of the equivalent widths, which is independent from the reddening, yields
$\rm H\alpha / H\beta > 12$. 
This is just about consistent with the values obtained from Williams'
radiative transfer models with temperatures around 8000\,K 
and $\log N_0 \approx 12$. 
Hence, the assumption of extremely high reddening does in principle allow
for the presence of a 
low temperature and low density accretion disc.
Such cool and low density discs however, are optically thin and thus
show strong emission lines in several neutral or easily excitable elements 
like He\,I, Ca\,II, and Fe\,II. No such lines
are present in the spectrum of V529\,Ori which makes the identification
as a CV and hence as an old nova doubtful.

The red continuum together with the
H$\alpha$ emission rather suggest that this object could be a classical 
T\,Tauri star or even a post T\,Tauri. Note, that we find some weak 
absorption around the Li\,I resonance line at 670.8\,nm but that our 
spectral resolution is too low for a clear assignment. High or medium 
resolved spectroscopy of this object is desired to clarify the presence
of Lithium and thus allow a proper classification of this object.
The FWHM of the H$\alpha$ line, which corresponds to a velocity of about 
700\,km/s, is high but still consistent with what is expected for
accretion on T\,Tauris (see e.g. Navascu\'es \& Mart\'{\i}n \cite{nava+03}). 

With these considerations and
taking into account the large uncertainty of the original coordinates,
we thus believe that the actual nova has still not been identified.

\section{CQ Velorum}
\begin{figure}
\rotatebox{-90}{\resizebox{!}{8.9cm}{\includegraphics{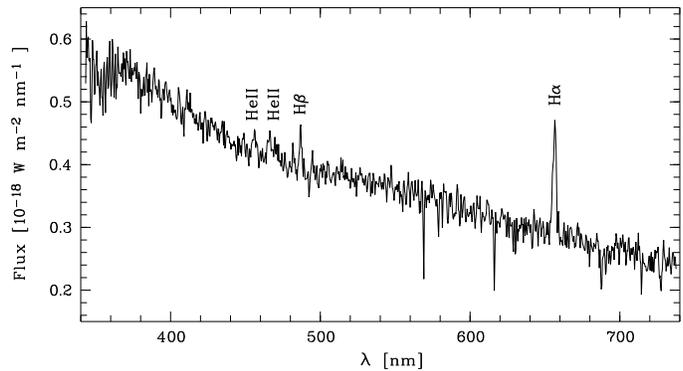}}}
\caption{\label{cqvels}Although the spectrum of CQ\,Vel is rather noisy,
the blue continuum together with the presence of Balmer and He\,II lines 
in emission confirm the identification of 
this object as old nova.}
\end{figure}

CQ\,Vel has been discovered in outburst by C.J. van Houten, Leiden, on 
Johannesburg plates from 1940 (van Houten \cite{hout50}). 
A follow--up investigation of Harvard plates revealed that the nova
reached $9.0^{\rm m}$ maximum light on April 19, 1940 
(Hoffleit \cite{hoff50}) and was a moderately
fast nova with $t_3 = 53$\,d (Duerbeck \cite{duer81}).
Hoffleit also indicates that the light curve 
showed strong brightness fluctuations, 
Borra \& Andersen (\cite{borr+70}) pointed out the similarity to
FH\,Ser (strong decline), Duerbeck (\cite{duer81})
classified the nova as type Cb (strong brightness decline during maximum),
and Rosenbusch (\cite{rose99}) compared its temporary fading to the 
light curve of DQ\,Her. All these comparisons or classifications 
refer to the same feature 
of the light curve, the sudden drop during the transition state, which is
best explained by assuming the production of dust in this state,
which would then cause the drop of brightness via extinction.
However, Harrison \& Gehrz (\cite{harr+94}) report that the system has not been
detected by IRAS, and thus 
the dust has either faded or cooled down significantly.

Using the 'Maximum magnitude versus rate of decline' method 
(Della Valle \& Livio \cite{dell+95}), Shafter (\cite{shaf97})
determined the distance of CQ\,Vel as $d = 9.3$\,kpc.
With this value and Rosenbusch's estimated nova shell radius of 0.2 mpc,
the apparent size of the shell computes to 0.01". Gill \& O'Brien 
(\cite{gill+98}) tried to map the nova remnant but found it to be not extended
at a seeing of 1.1". Munari \& Zwitter (\cite{muna+98}) tried to take a first
spectrum of the nova, but with $V \ge 21^{\rm m}$ the object was too faint
for their survey.

Duerbeck (\cite{duer87}) gives two possible candidates in the vicinity of the
nova. The high--speed photometry by Woudt \& Warner (\cite{woud+01})
revealed that the brighter of those candidates shows constant 
brightness, while the
slightly fainter one shows the typical CV flickering over a four hour
observing run and has hence been identified as the old nova.

The spectrum of CQ\,Vel is given in Fig.\ref{cqvels} and shows moderately 
strong Balmer emission as well as He\,II. Although of rather poor signal/noise,
it hence confirms Woudt \& Warner's identification.

\begin{figure*}
\rotatebox{-90}{\resizebox{!}{18.0cm}{\includegraphics{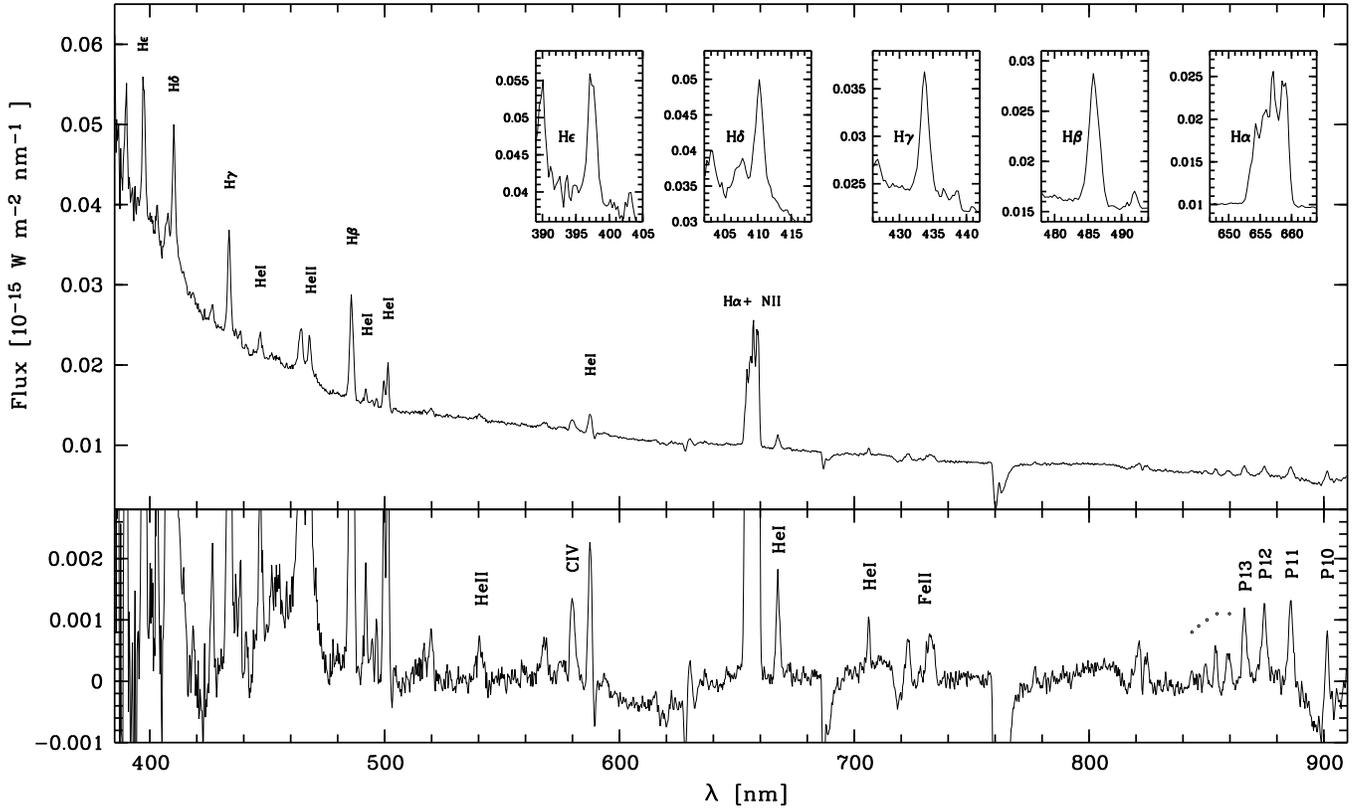}}}
\caption{\label{v842s}Spectrum of V842\,Cen. The upper plot shows the spectrum 
as observed. Notable are the blue continuum and the various strong emission lines
(hydrogen, He\,I, He\,II). The strongest Balmer lines have been
plotted individually to compare the line profiles; the contamination
of H$\alpha$ with N\,II becomes obvious. 
To check for the fainter lines, the continuum has 
been subtracted in the lower plot. Several weak, high excitation lines become 
visible, which probably originate in the hot nova shell, which is still present 
seventeen years after the explosion.}
\end{figure*}
\section{V842 Centauri}
The most recent nova of our sample is V842\,Cen, whose outburst in 1986
has been discovered by McNaught (\cite{mcna86}). He also identified the
nova progenitor as a faint (V$>$18) object on the UK Schmidt plates.
The nova reached a visual magnitude of 4.6 (Duerbeck \cite{duer87}). 
During its decline it has been well observed, both 
photometrically and spectroscopically. Whitelock (\cite{whit87})
determined $t_3 = 48$\,d and thus classified it as moderately fast nova. 
She also found evidence for dust formation in the optical and IR light curves.
The presence of dust is confirmed by infrared spectroscopy 
during and after the formation process 
(e.g. Wichmann et al. \cite{wich+90}; Smith et al. \cite{smit+94}).

H$\alpha$/[N{\sc ii}] observations by Gill \& O'Brien (\cite{gill+98}) 
revealed a shell of 1.6 arcsec diameter in 1995. In March 1998,
Downes \& Duerbeck (\cite{down+00}) took H$\alpha$ and [O{\sc iii}] images
which show an incomplete circular shell of $5.6 \times 6.0$\,arcsec diameter.
A single expansion velocity is not consistent with these two shell sizes.
However, emission line studies of the nova decline show two expansion velocity 
components, a high density region with $v_{\rm exp} =525\rm km\,s^{-1}$ and
low density material with $v_{\rm exp} = 2000\rm km\,s^{-1}$.
Downes \& Duerbeck showed that they can explain the two shell sizes by
assuming that their larger shell is built from the
low--density material with high expansion velocity while the
smaller shell of Gill \& O'Brien contains the slower expanding 
high--density material. This yields a distance of V842\,Cen of 
$\rm 1.3\pm 0.5 kpc$, which is consistent with values found from 
'maximum vs decline' methods and various reddening estimates 
(Sekiguchi et al. \cite{seki+89}).

Recent high speed photometry by Woudt \& Warner (\cite{woud+03}) show
a still active system, continuously
flaring on time scales of $\approx$5\,min, but no orbital modulation.
They thus conclude that the system is probably seen at low inclination.

The optical spectrum (see Fig. \ref{v842s}) 
is similar to that of XX\,Tau in the sense that the complete Balmer 
series is present in emission, and that He\,I is more dominant than He\,II.
In this respect, similar to XX\,Tau, the spectrum resembles that of a 
dwarf nova.
This case is especially interesting, as the outburst of V842\,Cen happened only
seventeen years ago, and the nova had not yet time to cool down.
This is supported by the extremely blue continuum. For the blue slope 
$F =  \lambda ^{\alpha}$ we derive $\alpha = -4.32(5)$ which is far bluer
than expected for a steady--state disc; 
for comparison, Lynden-Bell (\cite{lynd69}) calculates a slope
of $\lambda^{(-7/3)}$
for a large steady state disc radiating like a black body.
For $\lambda > 500$\,nm, the continuum
slope can instead be fitted with $\alpha = -1.59(2)$, this value being much 
more reasonable for an accretion disc.
The interpretation is hence that due to the short time 
the nova had to cool down, either the white dwarf itself or
a single region in the accretion disc are still extremely heated, thus
yielding an additional blue component to the continuum which is thus not
yet disc dominated.

\begin{table*}
\caption{\label{line_tab}
For each nova the outburst year, magnitude, and amplitude, the
slope $F = \lambda ^{\alpha}$ of the continuum, 
and the FWHM, equivalent widths, and line fluxes of the main emission lines
in the observed spectra are given. 
No de--reddening has been applied.
Note that the uncertainty of the line flux describes the
uncertainty of the relative flux in the line and does not include the
photometric error. Outburst magnitudes and amplitudes are taken/calculated from
the references as given in the text.
}
\begin{tabular}{l c c c c l c c c }
\hline
\hline
 \noalign{\smallskip}
Nova & Date & $m_{\rm max}$ & $\Delta m$ &$\alpha$ & Transition & FWHM [nm] & $-W$ [nm] & F [$10^{-18}$\,W\,m$^{-2}$] \\
 \noalign{\smallskip}
\hline
 \noalign{\smallskip}
V630\,Sgr & 1936 & ${\rm vis}=4.0^{\rm m}$ & 13.9$^{\rm m}$ &$-1.21 \pm 0.02^{1)}$ & H$\alpha$ & $2.36\pm 0.04$ & $2.70 \pm 0.06$ & $4.09\pm 0.09$ \\
         & & & & $-2.55 \pm 0.02^{1)}$ & H$\beta $ & $1.90\pm 0.04$ & $1.35 \pm 0.07$ & $3.32\pm 0.15$ \\
         & & & & & H$\gamma$ & $1.59\pm 0.04$ & $0.73 \pm 0.04$ & $2.15\pm 0.07$ \\
         & & & & & He\,II\,$\lambda$541 & $2.00\pm 0.09$ & $0.51\pm 0.03$ & $1.15\pm 0.06$ \\
         & & & & & He\,II\,$\lambda$469 & $1.93\pm 0.06$ & $2.8\pm 0.2$ & $7.0\pm 0.2$ \\
 \noalign{\smallskip}
\hline
 \noalign{\smallskip}
XX\,Tau & 1927 & $p=5.9^{\rm m}$ & $>13.1^{\rm m}$ & $-2.66\pm 0.03$ & H$\alpha$ &   $3.36 \pm 0.03$ & $5.2\pm 0.1$ & $1.56\pm 0.05$ \\
        & & & & & H$\beta $ &   $2.30 \pm 0.04$ & $1.92\pm 0.06$ & $1.29\pm 0.05$ \\
        & & & & & H$\gamma$ &   $2.21 \pm 0.05$ & $1.36\pm 0.03$ & $1.17\pm 0.01$ \\
        & & & & & H$\delta$ &   $2.56 \pm 0.04$ & $1.38\pm 0.1$  & $1.3\pm 0.1$   \\
        & & & & & H$\epsilon$ & $2.29 \pm 0.04$ & $1.11\pm 0.05$ & $1.12\pm 0.04$ \\
        & & & & & H$_8$ &       $2.05 \pm 0.07$ & $0.85\pm 0.05$ & $0.88\pm 0.04$ \\
        & & & & & H$_9$ &       $1.91 \pm 0.12$ & $0.53\pm 0.05$ & $0.58\pm 0.04$ \\
        & & & & & He\,I\,$\lambda$668 & $1.9 \pm 1.5$ & $0.39\pm 0.1$ & $0.11\pm 0.04$ \\
        & & & & & He\,I\,$\lambda$588 & $2.8 \pm 0.2$ & $0.50\pm 0.1$ & $0.22\pm 0.04$ \\
        & & & & & He\,I\,$\lambda$447 & $2.3 \pm 0.2$ & $0.35\pm 0.1$ & $0.31\pm 0.06$ \\
 \noalign{\smallskip}
\hline
 \noalign{\smallskip}
V529\,Ori & 1678 & ${\rm vis}=7^{\rm m}$ & & $+4.87\pm 0.07^{2)}$ &  H$\alpha$ & $1.43\pm 0.05$ & $2.95\pm 0.06$ & $3.75\pm 0.07$  \\
 \noalign{\smallskip}
\hline
 \noalign{\smallskip}
CQ\,Vel & 1940 & $p=9.0^{\rm m}$ & $>12.1^{\rm m}$ & $-1.15 \pm 0.03$ & H$\alpha$ & $2.88 \pm 0.04$ & $1.80\pm0.05$ & $0.52\pm 0.01$  \\
        & & & & & H$\beta $ & $2.1\pm 0.1$    & $0.45\pm0.05$ & $0.18\pm 0.01$  \\
 \noalign{\smallskip}
\hline
 \noalign{\smallskip}
V842\,Cen & 1986 & ${\rm vis}=4.6^{\rm m}$ & $>13.4^{\rm m}$ & $-4.32\pm 0.05 ^{3)}$ & H$\alpha$ & $5.08\pm 0.03$ & $7.31\pm 0.04$ & $0.0719\pm 0.0004$ \\
          & & & & $-1.59\pm 0.02 ^{3)}$ & H$\beta $ & $1.79\pm 0.02$ & $1.56\pm 0.03$ & $0.0248\pm 0.0003$ \\
          & & & & & H$\gamma$ & $1.5 \pm 0.05$ & $0.86\pm 0.02$ & $0.0208\pm 0.0003$ \\
          & & & & & H$\delta$ & $1.9 \pm 0.1$  & $0.8\pm 0.2$   & $0.027\pm 0.03$\\
          & & & & & H$\epsilon$&$1.42\pm 0.02$ & $0.67\pm 0.03$ & $0.0264\pm 0.0003$ \\
          & & & & & H$_8$     & $1.41\pm 0.03$ & $0.38\pm 0.05$ & $0.016\pm 0.001$  \\
          & & & & & He\,I\,$\lambda$706? & $1.95\pm 0.05$& $0.12\pm 0.02$ & $0.00105\pm 0.0003$ \\
          & & & & & He\,I\,$\lambda$668 & $1.86\pm 0.02$& $0.33\pm 0.02$ & $0.00314$\\
          & & & & & He\,I\,$\lambda$588 & $1.72\pm 0.02$& $>0.38$ & $>0.0044$\\
          & & & & & He\,I\,$\lambda$447 & $1.35\pm 0.2$ & $0.22\pm 0.02$ & $0.0045 \pm 0.0004$\\
          & & & & & C\,IV $\lambda$580 & $2.3\pm 0.2$& $0.30\pm 0.02$ & $0.0035 \pm 0.0004$\\
 \noalign{\smallskip}
\hline
\end{tabular}\\
$^1)$ The continuum of V630\,Sgr shows two slopes; the first for  
$\lambda < 582$\,nm, the second for $\lambda \ge 582$\,nm.\\
$^2)$ V529\,Ori shows no straight line in the double--logarithmic plot;
thus only a formal value for a linear fit to the data is given.\\
$^3)$ The continuum of V842\,Cen shows two slopes; the first for  
$\lambda < 500$\,nm, the second for $\lambda \ge 500$\,nm.
\end{table*}

Further evidence comes from the comparison of the line profile of H$\alpha$ 
with that of the other emission lines and especially the Balmer lines.
Its broad profile indicates that  H$\alpha$ is still strongly disturbed 
by N\,II and thus that V842\,Cen has not yet reached its quiescence level.
Also the presence of C\,IV at 580\,nm and several other high excitation lines
suggest that the hot nova shell is still present in the spectrum.
Therefore V842\,Cen is an even stronger case of a nova
that comprises a rather low mass transfer
rate and might have an accordingly short orbital period.

Except for H$\alpha$ and H$\delta$ which are obviously blended with other 
lines, the FWHM of the Balmer lines compute to an average velocity of
$1060\pm 30$\,km/s. For the lines of He\,I, instead, we find a significant 
lower average velocity of $790\pm 50$\,km/s. Most likely, these lines 
thus originate in different areas in the accretion disc. A detailed analysis
of time series spectroscopy is necessary to confirm this idea. The general low
values for the radial velocity component, 
as derived from the FWHM, confirm the conclusion that 
V842\,Cen is seen at low inclination.

\section{Discussion and Conclusion}
The spectroscopic analysis of old novae reveals a variety
of CV spectra not at all alike to each other 
(see e.g. Ringwald et al. \cite{ring+96}).
The hibernation scenario suggests that the older novae, 
which had already time to cool down, 
are more likely to be in a state of low mass transfer and are
thus less luminous than 'normal' novae and look like intermediate systems 
between novae and dwarf novae (e.g. Shara et al. \cite{shar+86}).
However, Ringwald et al. found no relation of the spectral appearance
to the age of the nova outburst.

In this context, the existence
of old novae like XX\,Tau and V842\,Cen 
is of special interest. Both novae had their outburst in the 20th century and
do as such not belong to the older novae. In fact, the outburst of V842\,Cen 
occurred very recently. Still, both novae show spectra which suggest them to 
be intermediate objects between novae and dwarf novae.
Although our spectra
are not conclusive regarding the hibernation hypothesis,  they are much more
dwarf-nova like than anything that has been presented before in favour of 
hibernation. 
Their existence therefore supports the idea that CVs undergo indeed 
evolution into different subtypes. 
So far, it is not possible to decide if this evolution 
happens in cycles as suggested by the hibernation model.

Taking into account the date of the outburst, the time elapsed 
since then is much less than predicted by hibernation for a nova to
appear like a dwarf nova. We therefore regard our
observations as a piece of evidence against hibernation in its current form.
Instead, since pre-novae tend to be of the same brightness or even fainter 
than post-novae 
(see e.g. Robinson \cite{robi75} and Retter \& Lipkin \cite{rett+01}) 
these two novae are actually likely to
originate from a CV subtype with rather low mass transfer rate, 
i.e. a dwarf nova.
The possible existence of such systems has recently been discussed by
Townsley \& Bildsten (\cite{town+04}).
Novae will of course have
different outburst probabilities depending on the time necessary for
accreting a sufficient amount of material onto the white dwarf 
and thus e.g. related to the mass transfer rate of the progenitor.
Using the  models of Townsley \& Bildsten (\cite{town+04}),
to estimate an average time $t_{\rm ign}$ for the accretion of the necessary 
material, one finds $t_{\rm ign} = 1000$\,yr for an accretion rate of
$\Delta M = 10^{-8} \rm M_{\odot}$, while $t_{\rm ign} = 10^8$\,yr are necessary
if the accretion rate is only $\Delta M = 10^{-11} \rm M_{\odot}$.
We therefore expect novae with nova--like origins to be far more numerous
than novae with dwarf--nova origin, which could explain why so few of these
systems are known.

To judge whether these low mass transfer novae look like dwarf novae 
because they originate
from dwarf novae or because they are on their way into hibernation, 
we need to know their orbital periods.
In the hibernation scenario
one would expect to find them at all periods, with a higher
appearance rate at longer periods where 'normal' novae are found. 
If instead the low mass transfer novae are found mostly at short orbital 
periods, i.e. below the period gap, this would rather indicate that the 
nova progenitors in these cases also are low mass transfer dwarf novae.
This does not
necessarily rule out the hibernation scenario, but 
it removes some observational evidence for it. 

To summarise, we would like to state that among the five, so far analysed TON 
candidates, four appear to be the rightly recovered novae, and two among these
four show spectroscopic evidence for being some intermediate 
object between a 
classical and a dwarf nova. However, due to their rather young age, their 
existence does not support the hibernation scenario. Further spectroscopy of
TONs are needed to get a larger sample of this nova type; the determination
of their orbital periods will help to determine their origin. 
Thus the remaining
statistical evidence for the hibernation scenario can be judged.

\acknowledgement{We acknowledge that this research has made intense 
use of the Simbad database operated at CDS, Strasbourg, France.}

\end{document}